\documentclass[prl,twocolumn,showpacs,preprintnumbers,amsmath,amssymb,superscriptaddress]{revtex4}
\usepackage{dcolumn}
\usepackage{bm}
\usepackage{graphicx}

\newcommand{\pa}{\partial}
\newcommand{\ep}{\varepsilon}

\begin{document}
\title{Geometry-Driven Shift in the Tomonaga-Luttinger Exponent of Deformed Cylinders}

\author{Hiroyuki Shima}
\email[Corresponding author: ]{shima@eng.hokudai.ac.jp}
\affiliation{Department of Applied Physics, Graduate School of Engineering,
Hokkaido University, Sapporo 060-8628, Japan}

\author{Hideo Yoshioka}
\affiliation{Department of Physics, Nara Women's University, Nara 630-8506, Japan}

\author{Jun Onoe}
\affiliation{Research Laboratory for Nuclear Reactors and Department of Nuclear Engineering,
Tokyo Institute of Technology, 2-12-1 Ookayama, Meguro, Tokyo 152-8550, Japan}

\date{\today}

%
%

\begin{abstract}
We demonstrate the effects of geometric perturbation on the Tomonaga-Luttinger liquid (TLL) states in a long, thin, hollow cylinder whose radius varies periodically. The variation in the surface curvature inherent to the system gives rise to a significant increase in the power-law exponent of the single-particle density of states. The increase in the TLL exponent is caused by a curvature-induced potential that attracts low-energy electrons to region that has large curvature.
\end{abstract}





\maketitle

Studying the quantum mechanics of a particle confined to curved surfaces 
has been a problem for more than fifty years. 
The difficulty arises from operator-ordering ambiguities \cite{DeWitt}, 
which permit multiple consistent quantizations for a curved system. 
The conventional method used to resolve the ambiguities is 
the confining-potential approach \cite{Jensen,daCosta}. 
In this approach, the motion of a particle on a curved surface
(or, more generally, a curved space) is regarded as being 
confined by a strong force acting normal to the surface. 
Because of the confinement, quantum excitation energies 
in the normal direction are raised far beyond those 
in the tangential direction. 
Hence, we can safely ignore the particle motion normal to 
the surface, which leads to an effective Hamiltonian 
for propagation along the curved surface with no ambiguity.

It is well known that the effective Hamiltonian involves 
an effective scalar potential whose magnitude depends on 
the local surface curvature \cite{Jensen,daCosta,Kaplan,Jaffe}. 
As a result, quantum particles confined to a thin, curved layer 
behave differently from those on a flat plane, even in 
the absence of any external field (except for the confining force). 
Such curvature effects have gained renewed attention in the last decade, 
mainly because of the technological progress that has enabled 
the fabrication of low-dimensional nanostructures with complex 
geometry\cite{exp1,exp2,exp3,exp4,exp5,exp6,ShimaNTN,exp7,Arroyo}. 
From the theoretical perspective, many intriguing phenomena 
pertinent to electronic 
states \cite{state1,state2,state3,state4,state5,state6,state7,state8}, 
electron diffusion \cite{diff}, 
and electron transport \cite{trans1,trans2,trans3,trans4} have been suggested. 
In particular, the correlation between surface curvature and 
spin-orbit interaction \cite{spin1,spin2} as well as with 
the external magnetic field \cite{mag1,mag2,mag3} has been recently 
considered as a fascinating subject. 

Most of the previous works focused on noninteracting electron systems, 
though few have focused on interacting electrons \cite{Interaction} and 
their collective excitations. 
However, in a low-dimensional system, Coulombic interactions 
may drastically change the quantum nature of the system. 
Particularly noteworthy are one-dimensional systems, 
where the Fermi-liquid theory breaks down so that the system is 
in a Tomonaga-Luttinger liquid (TLL) state \cite{TL1}. 
In a TLL state, many physical quantities exhibit 
a power-law dependence stemming from the absence of 
single-particle excitations near the Fermi energy; 
this situation naturally raises the question as to 
how geometric perturbation affects the TLL behaviors of 
quasi one-dimensional curved systems. 
Peanut-shaped C$_{60}$ polymers \cite{Onoe,Toda}
and MoS$_2$ hollow nanotubes \cite{MoS2} are exemplary materials 
to be considered for studying TLL behaviors; they are thin, 
long, hollow tubules whose radius is periodically modulated 
along the tube axis. 
Hence, the periodic surface curvature intrinsic to the systems 
will produce sizeable effects on their TLL properties 
if they exhibit one-dimensional metallic properties.

In this Letter, we examined the effects of geometric curvature on the
TLL states in quantum hollow cylinders
with a periodically varying radius and demonstrate that 
the presence of a curvature-induced potential can yield
a significant increase in the power-law exponent $\alpha$
of the single-particle density of states $n(\omega)$
near the Fermi energy $E_F$; {\it i.e.,}
$n(\omega) \propto |\hbar\omega - E_F|^{\alpha}$ \cite{Yoshi1,Yoshi2}.
The geometric conditions required for the shift in $\alpha$ to be observable
are within the realm of laboratory experiments,
which implies that our predictions can be verified with existing materials.

We first considered noninteracting spinless electrons
confined to a general two-dimensional curved surface $S$ embedded in
a three-dimensional Euclidean space. A point $\bm{p}$ on $S$
is represented by
$\bm{p} = (x(u^1,u^2), y(u^1,u^2), z(u^1,u^2))$,
where $(u^1, u^2)$ is a curvilinear coordinate spanning the surface
and $(x,y,z)$ are the Cartesian coordinates in the embedding space.
Using the notation $\bm{p}_i \equiv \pa \bm{p}/\pa u^i$ $(i=1,2)$,
we introduced the following quantities:
$g_{ij} = \bm{p}_i \cdot \bm{p}_j$, 
$h_{ij} = \bm{p}_{ij} \cdot \bm{n}$,
$\bm{n} = (\bm{p}_i \times \bm{p}_j)/\|\bm{p}_i \times \bm{p}_j\|$,
where $\bm{n}$ is the unit vector normal to the surface.
Using the confining-potential approach \cite{Jensen,daCosta},
we obtained the Schr\"odinger equation for noninteracting electron
systems on curved surfaces as follows:
\begin{equation}
-\frac{\hbar^2}{2m^*}
\left[
\frac{1}{\sqrt{g}} \sum_{i,j=1}^2 \frac{\pa}{\pa u^i} \sqrt{g}
g^{ij} \frac{\pa}{\pa u^j}
+({\cal H}^2 - {\cal K})
\right]
\Psi
=
E \Psi,
\label{eq_002}
\end{equation}
where  $g = {\rm det} (g_{ij})$, $g^{ij} = g_{ij}^{-1}$ \cite{Higher},
and $m^*$ is the effective mass of electrons.
The quantities
${\cal K} = (h_{11}h_{22}-h_{12}^2)/g$
and
${\cal H} = (g_{11}h_{22} + g_{22} h_{11} - 2g_{12} h_{12})/(2g)$
are the so-called Gaussian curvature and mean curvature, respectively,
both of which are functions of $(u^1, u^2)$.
The term ${\cal H}^2 - {\cal K}$ in Eq.~(\ref{eq_002})
is the effective scalar potential induced by surface curvature.

\begin{figure}[ttt]
\includegraphics[scale=0.37]{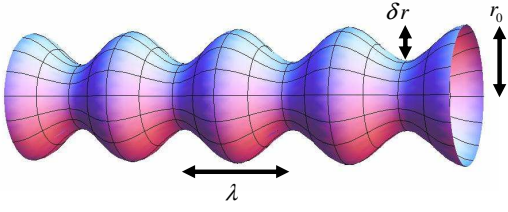}
\caption{(color online) Schematic illustration of a quantum hollow cylinder 
with periodic radius modulation.}
\label{fig_00}
\end{figure}

We next focused on a hollow tube with a periodically varying radius
represented by
$\bm{p} = (r(z)\cos\theta, r(z)\sin\theta, z)$ (see Fig.~\ref{fig_00}).
The tube radius $r(z)$ 
is periodically modulated in the axial $z$ direction as
\begin{equation}
r(z) = r_0 - \frac{\delta r}{2} + \frac{\delta r}{2} \cos \left( \frac{2\pi}{\lambda} z \right),
\label{eq_003}
\end{equation}
where the parameters $r_0$ and $\delta r$ are introduced to express
the maximum and minimum of $r(z)$ as
$r_0$ and $r_0-\delta r$, respectively.
Because of the rotational symmetry, the eigenfunctions of the system
have the form of
$\Psi(z,\theta) = e^{in\theta} \psi_n(z)$.
Thus, the problem reduces to the one-dimensional
Schr\"odinger equation
\begin{equation}
-\frac{\hbar^2}{2m^*} 
\left[
\frac{1}{r f}
\frac{d}{dz}
\frac{r}{f}
\frac{d}{dz}
-
\frac{n^2}{r^2}
+ \!
({\cal H}^2 - {\cal K})
\right]
\psi_n(z)
=
E \psi_n (z),
\label{eq_005}
\end{equation}
where  $f(z) = \sqrt{1+{r'}^2}$,
${\cal K} = -r''/(r f^2)$,
and
${\cal H} = (f^2 - rr'' )/(2rf^3)$
with $r'\equiv dr/dz$.

\begin{figure}[ttt]
\hspace{-5mm}
\includegraphics[scale=0.38]{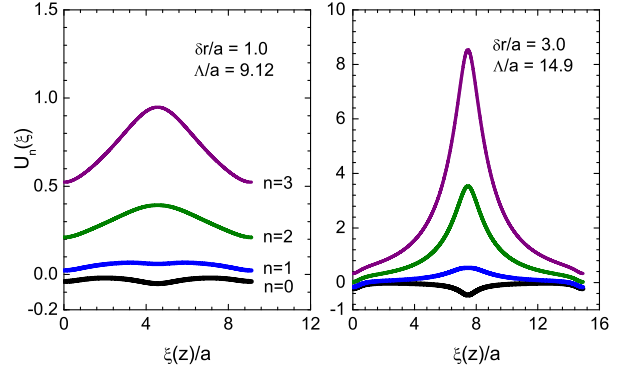}
\caption{(color online) Profiles of the curvature-induced effective potential $U_n(\xi)$
for one period $[0,\Lambda]$.
Geometric parameters $r_0 = 4.0$ and $\lambda = 8.0$ in units of $a$
are fixed.
Integers $n$ represent the angular momentum of eigenstates
in the circumferential direction of a hollow tube.}
\label{fig_01}
\end{figure}

Equation (\ref{eq_005}) is simplified by 
using a new variable 
$\xi = \xi(z) = \int_0^z f(\eta) d\eta$,
which corresponds to the line length
along the curve on the surface with a fixed $\theta$.
Straightforward calculation yields \cite{state5}
\begin{eqnarray}
\!\!\!\!\!\!\!\!\!\!\!\!\!\!\!\!\!\!& &\left[
- a^2 \frac{d^2}{d \xi^2} + U_n(\xi)
\right] \psi_n(\xi) = \ep \psi_n(\xi),
\;\;
\ep = \frac{2m^* a^2 E}{\hbar^2}
\label{eq_010}
\end{eqnarray}
with
$U_n (\xi) = (n^2 - \frac14) a^2/r^2 - {r''}^2 a^2/(4 f^6)$,
where $r$, $r''$, and $f$ are regarded as the functions of $\xi$ 
using the inverse relation $z = z^{-1}(\xi)$.
In order to derive Eq.~(\ref{eq_010}), we introduced the length scale, $a$,
and then multiplied both sides of Eq.~(\ref{eq_005}) with $a$
to make the units of $U_n$ and $\ep$ dimensionless.
Notice that by the definition of $\xi(z)$,
$U_n$ is periodic with a period $\Lambda = \xi(\lambda)$
depending on $r_0$ and $\delta r$ (as well as $\lambda$).

Figure \ref{fig_01} shows the spatial profile of $U_n$ within one period;
throughout the present work, we fixed $r_0 = 4.0$ and $\lambda = 8.0$ in units of $a$
by simulating the geometry of actual peanut-shaped C$_{60}$ polymers
whose geometry is reproduced by imposing $a=1$ \AA.
We found that $U_n$ takes extrema at $\xi=0$ (or $\Lambda$) and $\xi=\Lambda/2$, 
where $r$ takes the maximum $(r=r_0)$ and the minimum $(r=r_0-\delta r)$ values,
respectively.
It also follows that $U_n$ for $n=0$ is negative for any $\xi$
as derived from the definition of $U_n$.

\begin{figure}[ttt]
\hspace{-5mm}
\vspace{0mm}
\includegraphics[scale=0.40]{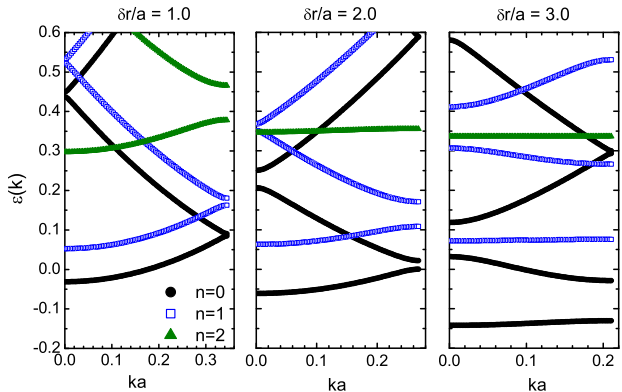}
\caption{(color online) Energy-band structure of a sinusoidal hollow tubule with
$r_0 = 4.0$ and $\lambda = 8.0$ in units of $a$.
}
\label{fig_02}
\end{figure}

To solve Eq.~(\ref{eq_010}),
we use the Fourier series expansions
$U_n(\xi) = \sum_G U_G^{(n)} e^{iG \xi}$
and
$\psi_n(\xi) = \sum_k c_k^{(n)} e^{ik\xi}$,
where $G = \pm 2\pi j/ \Lambda$ $(j=0,1,2,\cdots)$.
Substituting the expansions into Eq.~(\ref{eq_010}), we obtain
the secular equation
$( \frac{\hbar^2}{2m^*} k^2 -\ep ) c_k^{(n)} 
+
\sum_{G=-G_c}^{G_c} U_G^{(n)} c_{k-G}^{(n)} = 0$
that holds for all possible $k$s and $n$s.
The summation has been truncated by $G_c = 20\pi/\Lambda$
because of the rapid decay of $U_G$ with $|G|$.
We then numerically calculated
eigenvalues for $0\le k \le \pi/\Lambda$
and evaluated the low-energy band structure
for several different $\delta r$, as depicted in Fig.~\ref{fig_02}.
In all the cases, there is some energy gap at 
the Brillouin zone boundary $G_0 \equiv \pi/\Lambda$,
where a wider energy gap occurs for a larger $\delta r$,
as expected from the large amplitude of $|U_n(\xi)|$
with increasing $\delta r$
(see Fig.~\ref{fig_01}).

We now consider the Coulombic interactions between spinless electrons.
The interactions make the electron-hole pairs share
the ground state of the noninteracting electron system, wherein 
the most strongly affected states are those lying 
in the vicinity of $E_F$.
As a consequence,
the single-particle density of states $n(\omega)$
near $E_F$ exhibits a power-law singularity of the following form \cite{TL1}
\begin{equation}
n(\omega)\propto |\hbar \omega - E_F|^{\alpha},
\quad \alpha = \frac{K+K^{-1}}{2} -1.
\label{eq_025}
\end{equation}
The explicit form of $K$ is derived using the bosonization procedure \cite{TL1}
as follows
\begin{equation}
K = \lim_{q\to 0}
\sqrt{ \frac{2\pi \hbar v_F + g_4(q) - g_2(q)}{2\pi \hbar v_F + g_4(q) + g_2(q)} }.
\label{eq_030}
\end{equation}
Here, $v_F = \hbar^{-1} dE/dk|_{k=k_F}$ is the Fermi velocity,
and $g_4(q) = V(q,m)$ and $g_2(q) = V(q,m) - V(2k_F,m)$ are 
$q$-dependent coupling constants.
$V(q,m)$ is the Fourier transform of 
the screened interaction $V(\bm{r}) = -{\rm e}^2 e^{-\kappa |\bm{r}|}/(4\pi \ep |\bm{r}|)$,
where $\ep$is the dielectric constant and $\kappa$ is the screening length.
The transformation is performed in terms of the curvilinear coordinates $(\xi,\theta)$,
and thus, the resulting $V(q,m)$ becomes a function 
of both the momentum and angular momentum transfers,
$q$ and $m$, respectively.
To make concise arguments, $E_F$ was assumed to lie
in the lowest energy band $(n=0)$.
This allows us to eliminate the index $m$ from $V(q,m)$,
which leads to $V(q) = - \frac{{\rm e}^2}{4\pi \varepsilon}
\log [ ( q^2 + \kappa^2 ) r_0^2 ]$
for $q r_0 \ll 1$,
in which $r_0$ serves as the short-length-scale cut-off.

The aim of the present study is to examine the
$\delta r$-dependence of $K$ and $\alpha$,
which requires quantification of $k_F$ and $v_F$.
Among the many alternatives, we set $k_F a = 2.0\ell \times 10^{-2}$ $(\ell=1,\cdots,8)$,
and evaluate the dimensionless Fermi velocity defined by
$\tilde{v}_F \equiv d\ep/d(ka)|_{k=k_F}$ for each $k_F$;
the original $v_F$ is recovered by the relation $v_F = \hbar \tilde{v}_F/(2m^* a)$.
We chose the maximum value of $k_F$ such that it does not to exceed 
the $\delta r$-dependent zone boundary $G_0$ for all $\delta r$;
it readily follows that $G_0$ takes the minimum value at $\delta r = r_0$,
resulting in $G_0 a \simeq 0.17$,
and hence, $k_F < G_0$ can be satisfied for all $\delta r$.

\begin{figure}[ttt]
\hspace{-10mm}
\vspace{0mm}
\includegraphics[scale=0.42]{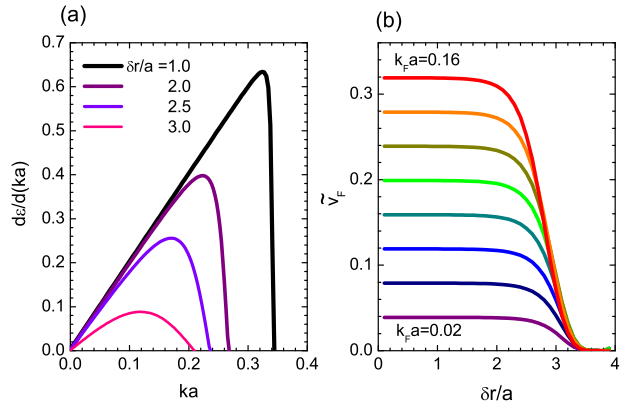}
\caption{(color online) (a) $k$-dependence of the derivative $d\ep/dk$ 
for the lowest energy band.
(b) $\delta r$-dependence of the dimensionless Fermi velocity 
$\tilde{v}_F = d\ep/d(ka)|_{k=k_F}$ for different $k_F a$
with increments $\Delta (k_F a) = 0.02$.}
\label{fig_03}
\end{figure}

Figure \ref{fig_03} (a) shows the plot of $d\epsilon/dk$ as a function of $ka$,
from which $\tilde{v}_F$ for various $k_F$ and $\delta r$ are deduced,
as shown in Fig.~\ref{fig_03} (b).
Figure \ref{fig_03} (b) shows that 
$\tilde{v}_F$ for all $k_F$ remains constant for $\delta r/a < 2.0$,
but decreases remarkably with increasing $\delta r$
for $\delta r/a > 2.5$.
The decrease in $\tilde{v}_F$ is caused by
the decrease in the slope of dispersion curves $\ep = \ep(k)$
with $\delta r$ (see Fig.~\ref{fig_02}).
The results shown in Fig.~\ref{fig_03} (b) imply that a change in $\delta r$
leads to a quantitative alteration in $K$ and $\alpha$,
particularly at $\delta r/a > 2.5$, and furthermore,
the degree of alteration increases for larger $k_F$,
as will be demonstrated later.

\begin{figure}[ttt]
\hspace{-15mm}
\vspace{0mm}
\includegraphics[scale=0.41]{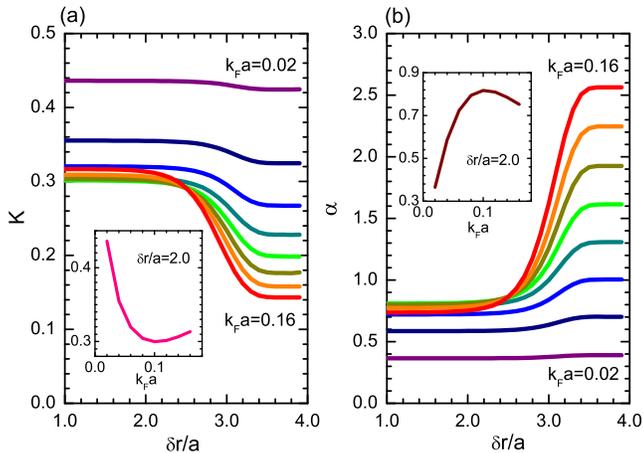}
\caption{(color online) $\delta r$-dependences
of $K$ and $\alpha$
defined by Eqs.~(\ref{eq_025}) and (\ref{eq_030}), respectively.
The screening parameter $\kappa$ is set to be 
$\kappa a = 1.0\times 10^{-3}$.
Insets: Nonmonotonic behaviors of $K$ and $\alpha$
as a function of $k_F$ at $\delta r/a = 2.0$.
}
\label{fig_04}
\end{figure}

Figure \ref{fig_04} shows the $\delta r$-dependence of both $K$ and $\alpha$
for different $k_F$ values.
According to the bosonization procedure \cite{TL1},
we set the screening parameter $\kappa$ to be 
$\kappa a = 1.0\times 10^{-3}$, which is smaller than
all $k_F a$ values that we have chosen.
We also set the interaction-energy scale
${\rm e}^2/(4\pi \ep a)$ to be $1.1$
in units of $\hbar^2/(2 m^* a^2)$
by simulating that of C$_{60}$-related materials \cite{ep,mass}.
The insets in Fig.~\ref{fig_04} shows the $k_F$-dependence of $K$ and $\alpha$
at $\delta r/a=2.0$; each $K$ and $\alpha$ take the minimum and maximum values,
respectively,
at the specific $k_F$ satisfying the relation 
$(d/dk_F) \log[(2\pi \hbar v_F + g_4)/g_2] = 0$,
which is equivalent to $dK/dk_F = 0$.
This result is the same for all $\delta r$ at $\delta r/a<2.0$.

The salient features of Fig.~\ref{fig_04} are the
significant decrease in $K$ and increase in $\alpha$ with 
an increase in $\delta r$ for $\delta r/a > 2.5$, as predicted earlier.
It is interesting to note that such $\delta r$-driven shifts in $K$ and $\alpha$
are attributed to the effects of geometric curvature on the nature of TLL states.
In fact, an increase in $\delta r$ 
amplifies the curvature-induced effective potential $U_n(\xi)$,
thus yielding a monotonic decrease in $\tilde{v}_F$
at $\delta r/a > 2.5$ (see Fig.~\ref{fig_03}).
The decrease in $\tilde{v}_F$ plays a dominant role
in the numerator of the expression in Eq.~(\ref{eq_030}),
and eventually leads to the systematic shifts in $K$ and $\alpha$.
We have confirmed that a change in the value of $\kappa$
does not substantially affect the behaviors of $K$ and $\alpha$ 
in a qualitative sense,
while the absolute values of $K$ and $\alpha$
moderately depend on the choice of $\kappa$.

An important consequence of the results shown in Fig.~\ref{fig_04}
is that nonzero surface curvature 
yields diverse alterations in the TLL behaviors of deformed cylinders.
This is because various kinds of power-law exponents 
observed in TLL states are related to the quantity $K$;
some such exponents are the exponent of power-law decay 
in Friedel oscillation \cite{TL3} 
and that of temperature (or voltage)-dependent conductance \cite{TL4}.
The present theoretical predictions need to be confirmed experimentally, 
thus opening a new field of science that deals with 
quantum electron systems on curved surfaces.

In conclusion, we reveal that 
the power-law exponent $\alpha$ of the TLL states
in deformed hollow nanocylinders
shows a monotonic increase with
an increase in the degree of surface curvature.
The increase in $\alpha$ is attributed to 
the curvature-driven effective potential $U_n(\xi)$
that acts on electrons moving along the curved surface. 
The present results suggest that there are shifts in the power-law exponents of TLL states
of real low-dimensional materials such as the peanut-shaped
C$_{60}$ polymers and MoS$_2$.

\acknowledgments
We acknowledge Prof.~K.~Yakubo, Prof.~Y.~Toda, Dr.~A.~Tokuno and Dr.~T.~Nishii
for stimulating discussions.
This study was supported by a Grant-in-Aid for Scientific
Research from the MEXT, Japan.
One of the authors (H.S.) is thankful for the financial support from 
Executive Office of Research Strategy in Hokkaido University.
A part of numerical simulations were carried out using
the facilities of the Supercomputer Center, ISSP, University of Tokyo.

%
%


\begin{thebibliography}{99}



\bibitem{DeWitt} B. De Witt, Phys.~Rev. {\bf 85} (1952) 635; 
Rev.~Mod.~Phys. {\bf 29} (1957) 377.
\bibitem{Jensen} H.~Jensen and H.~Koppe, Ann.~of Phys. {\bf 63} (1971) 586.
\bibitem{daCosta} R.~C.~T.~da Costa, Phys.~Rev.~A {\bf 23} (1982) 1981.
\bibitem{Kaplan} L.~Kaplan, N.~T.~Maitra and E.~J.~Heller, 
Phys.~Rev.~A. {\bf 56} (1997) 2592.
\bibitem{Jaffe} P.~C.~Schuster and R.~L.~Jaffe, Ann.~Phys. {\bf 307} (2003) 132.
\bibitem{exp1} O.~G.~Schmidt and K.~Eberl, Nature {\bf 410} (2001) 168.
\bibitem{exp2} S.~Tanda {\it et al.,}
Nature {\bf 417} (2002) 397.
\bibitem{exp3} Y.~Oshima, A.~Onga and K.~Takayanagi, 
Phys.~Rev.~Lett. {\bf 91} (2003) 205503.
\bibitem{exp4} A.~Lorke, S.~Bohm and W.~Wegscheider, 
Superlattices Microstruct. {\bf 33} (2003) 347.
\bibitem{exp5} D.~N.~McIlroy {\it et al.,}
J. Phys.: Condens. Matter {\bf 16} (2004) R415, and references therein.
\bibitem{exp6} M.~Sano {\it et al.,}
Science {\bf 293} (2004) 1299.
\bibitem{ShimaNTN} H.~Shima and M.~Sato, Nanotechnology {\bf 19} (2008) 495705;
phys.~stat.~solidi (b) {\it in press}.
\bibitem{exp7} T.~Fujita {\it et al.,}
Appl.~Phys.~Lett. {\bf 92} (2008) 251902.
\bibitem{Arroyo}
I.~Arias and M.~Arroyo, Phys.~Rev.~Lett. {\bf 100} (2008) 085503.
\bibitem{state1} A.~Mostafazadeh, Phys.~Rev.~A. {\bf 54} (1996) 1165.
\bibitem{state2} G.~Cantele, D.~Ninno and G.~Iadonisi, Phys.~Rev.~B. {\bf 61} (2000) 13730.
\bibitem{state3} H.~Aoki {\it et al.,}
Phys.~Rev.~B {\bf 65} (2001) 035102;
M.~Koshino and H.~Aoki, Phys.~Rev.~B {\bf 71} (2005) 073405.
\bibitem{state4} M.~V.~Entin and L.~I.~Magarill, Phys.~Rev.~B. {\bf 66} (2002) 205308.
\bibitem{state5} N.~Fujita, J.~Phys.~Soc.~Jpn. {\bf 73} (2004) 3115.
\bibitem{state6} 
J.~Gravesen and M.~Willatzen, Phys.~Rev.~A {\bf 72} (2005) 032108.
\bibitem{state7} H.~Taira and H.~Shima, Surf.~Sci. {\bf 601} (2007) 5270.
\bibitem{state8} V.~Atanasova and R.~Dandoloff, Phys.~Lett.~A {\bf 372} (2008) 6141.
\bibitem{diff} S.~K.~Baek, S.~D.~Yi and B.~J.~Kim, Phys.~Rev.~E {\bf 77} (2008) 022104.
\bibitem{trans1} D.~V.~Bulaev, V.~A.~Geyler, and V.~A.~Margulis, 
Phys.~Rev.~B {\bf 69} (2004) 195313.
\bibitem{trans2} A.~V.~Chaplik and R.~H.~Blick, New J.~Phys. {\bf 6} (2004) 33.
\bibitem{trans3} A.~Marchi {\it et al.,}
Phys.~Rev.~B {\bf 72} (2005) 035403.
\bibitem{trans4} G.~Cuoghi, G.~Ferrari and A.~Bertoni, Phys.~Rev.~B {\bf 79} (2009) 073410.
\bibitem{spin1} M.~V.~Entin and L.~I.~Magarill, Phys.~Rev.~B. {\bf 64} (2001) 085330;
Europhys.~Lett. {\bf 68} (2004) 853.
\bibitem{spin2} E.~Zhang, S.~Zhang and Q.~Wang, Phys.~Rev.~B {\bf 75} (2007) 085308.
\bibitem{mag1} M.~Encinosa, Phys.~Rev.~A {\bf 73} (2006) 012102.
\bibitem{mag2} O.~Olendski and L.~Mikhailovska, Phys.~Rev.~B {\bf 72} (2005) 235314;
{\it ibid.,} {\bf 77} (2008) 174405.
\bibitem{mag3} G.~Ferrari and G.~Cuoghi, Phys.~Rev.~Lett. {\bf 100} (2008) 230403;
G.~Ferrari {\it et al}., Phys.~Rev.~B {\bf 78} (2008) 115326.
\bibitem{Interaction} G.~Parascandolo {\it et al.,}
Phys.~Rev.~B. {\bf 68} (2003) 245318.
\bibitem{TL1} J.~Voit, Rep.~Prog.~Phys. {\bf 57} (1994) 977.
\bibitem{Onoe} J.~Onoe {\it et al.,}
Appl.~Phys.~Lett. {\bf 82} (2003) 595;
J.~Onoe {\it et al.,}
Phys.~Rev.~B {\bf 75} (2007) 233410.
\bibitem{Toda} Y.~Toda, S.~Ryuzaki and J.~Onoe, Appl.~Phys.~Lett. {\bf 92} (2008) 094102.
\bibitem{MoS2} P.~Santiago {\it et al.,}
Appl.~Phys.~A {\bf 78} (2004) 513.
\bibitem{Yoshi1} H.~Yoshioka, Physica E {\bf 18} (2003) 212.
\bibitem{Yoshi2} H. Ishii {\it et al}., 
Nature {\bf 426} (2003) 540.
\bibitem{Higher} H.~Shima and T.~Nakayama, 
{\it Higher Mathematics for Physics and Engineering},
(2009) (Springer-Verlag)
\bibitem{TL3} R.~Egger and H.~Grabert, Phys.~Rev.~Lett. {\bf 75} (1995) 3505.
\bibitem{TL4} C.~L.~Kane and M.~P.~A.~Fisher, Phys.~Rev.~B {\bf 46} (1992) 15233.
\bibitem{ep} A.~F.~Hebard {\it et al.,}
Appl.~Phys.~Lett. {\bf 59} (1991) 2109.
\bibitem{mass} A.~Oshiyama {\it et al.,}
J.~Phys.~Chem.~Solid., {\bf 53} (1992) 1457.


\end{thebibliography}
\end{document}